\documentclass{aastex6}
\PassOptionsToPackage{linktocpage}{hyperref}
\usepackage{amsmath}

\begin{document}


\title{Magnetoacoustic and Alfv\'{e}nic Black Holes
 with Hawking Radiation at Horizons Made of Magnephonons and Alphonons}



\author{Akbar Gheibi \altaffilmark{1} }

\author{Hossein Safari \altaffilmark{2} }
\affil{Department of Physics, University of Zanjan, P. O. Box 45195-313, Zanjan, Iran}

\author{Davina E. Innes\altaffilmark{3}}
\affil{Max-Planck Institute for Solar System Research, 37077 G\"{o}ttingen, Germany}

\and





\begin{abstract}

We introduce analogue black holes (BHs) based on ideal magnetohydrodynamic equations. Similar to acoustic BHs, which trap phonons and emit Hawking radiation (HR) at the sonic horizon where the flow speed changes from super- to sub-sonic, in the horizon of magnetoacoustic and Alfv\'{e}nic BHs, the magnetoacoustic and Alfv\'{e}n waves will be trapped and emit HR made of quantized vibrations similar to phonons which we call magnephonons and Alphonons. We proposed that magnetoacoustic and Alfv\'{e}nic BHs may be created in the laboratory using a tube with variable cross section embedded in a uniform magnetic field, and a super-magnetoacoustic or a super-Alfv\'{e}nic flow. We show that the Hawking temperature for both BHs is a function of the background magnetic field, number density of fluid, and radius of the tube. For a typical setup, the temperature is estimated to be about 0.0266 K.

\end{abstract}




\section{Introduction} \label{sec:intro}

In 1916 Schwarzschild gave a metric as a solution of the Einstein field equation. Singularity of such a metric predicted a gravitational black hole (BH) with event horizon at Schwarzschild radius \citep[see][]{Misner}. Although, based on classical physics everything, even light, is absorbed by BHs and cannot escape, in the context of quantum field theory in curved space, Hawking showed that BHs should emit black body radiation \citep{2,3}. Hawking radiation (HR) in the universe has not been observed yet, but numerous attempts have been done to simulate the interesting phenomena in the laboratory. Unruh showed that HR is not only a characteristic of gravitational BHs, it is also a characteristic of the acoustic analogue BH \citep{4,5}. After 1981, most of attempts are proposed based on Bose-Einstein condensates of quantum fluid \cite[]{6,7,8,9,10}, quasi particles in superfluid \citep{11}, ultra-cold fermions \citep{12}, in plasmas and ion rings \cite[]{13,14,15,16,17}, slow light in an atomic vapor \cite[]{18,19,20,21}, in water \citep{22,23},  etc. Recently, observation of self-amplifying HR in an analogue BH laser suggested a very promising experiment method for probing the inside of a BH \citep{24}.
 From a theoretical point of view, the acoustic analogue BH models are developed in geometrical acoustics and physical acoustics \citep{25}. Using the linearized hydrodynamic equations in the presence of initial material flow, a wave equation for velocity potential was obtained. Tensorial form of the wave equation results in an acoustic metric. The acoustic metric is singular at a point where the local sound speed is equal to the flow speed \citep{26}. This was interpreted as characteristic of a sonic BH \citep{25,26}. The effect of magnetic field on the acoustic BH and HR has not been studied, yet.
 Alternatively, the idea for definition of acoustic BH can be applied to
 introduce new analogue magnetoacoustic and Alfv\'{e}nic BHs in the magnetohydrodynamics  (MHD)
  framework.      	

Magnetohydrodynamics  (MHD) is an useful approach to analyze characteristics (flow, wave and dissipation, etc) of the  laboratory and astrophysical
plasma (e.g., \citep{Goedbloe}).
After 1942 \citep{Alf}, MHD  (Alfv\'{e}n and magnetoacoustic) waves have been detected, using the laboratory experiments  (e.g., \citep{Lundquist,Bostick, Lehnert}), in the earth-ionosphere and magnetic field \citep{Berthold,Masahisa}, a variety normal modes of the solar corona \cite[]{Aschwanden,Nakariakov,Wang,Aschwanden1,McIntosh}.
 Here, first we treat the magnetoacoustic, Alfv\'{e}n, and acoustic waves based on Helmholtz theorem for a uniform and stationary medium with constant background magnetic field. Second, mimicking the definition of acoustic BH in a nozzle, we introduce magnetoacoustic and Alfv\'{e}nic BHs with using a slightly variable cross section tube. We conclude that at the horizon of magnetoacoustic and Alfv\'{e}nic BHs  should emit radiations made of the  Magnephonon and Alphonon, respectively. We define two quasi-particles Magnephonon and Alphonon correspond to quantum of the magnetoacoustic and Alfv\'{e}n waves, respectively.

This paper is organized as follows: Sec. \ref{sec:metric} gives
 the properties of magnetoacoustic waves (fast, slow, and Alfv\'{e}n waves) using the Helmholtz decomposition and
 explains a derivation
of magnetoacoustic metric in the basis
of linearized ideal magnetohydrodynamic equations. Sec. \ref{sec: bhs}
introduces the acoustic, magnetoacoustic, and Alfv\'{e}nic black holes,
 respectively.
  Sec. \ref{sec: HT} calculate the Hawking temperature for
 the acoustic, magnetoacoustic, and Alfv\'{e}nic black holes. Sec. \ref{sec: conc}
 describes the conclusions.

\section{ Magnetoacoustic  waves and metric}\label{sec:metric}
Here, we give the conditions for a definition of a magnetoacoustic metric,  in the non-relativistic magnetohydrodynamic (MHD) framework. In the MHD approach, the behavior of continuous plasma is governed by a non-relativistic form of Maxwell's equations, together with Ohm's law, a gas law, equation of mass continuity, motion and energy equations \citep{27}. The ideal MHD equations for an adiabatic process and irrotational flow  $\nabla\times{\bf
v}=0$ are given by:
\begin{eqnarray}
&&\frac{{\partial \rho }} {{\partial t}} + \nabla.(\rho {\bf v})=
0,\label{cont}\\
 && \rho\left(\frac{\partial {\bf v} }{{\partial t}}
+\frac{1}{2}\nabla \text v^2\right) = -\nabla p +\frac
{1}{\mu}(\nabla\times {\bf B})\times{\bf B},\label{mom}\\
&&\frac{\partial{\bf B}}{\partial t}=\nabla\times({\bf
v}\times{\bf B}),\label{induc}\\
&&\nabla.{\bf B}=0,\label{solo}\\
&&p=k\rho^\gamma,\label{adia}
\end{eqnarray}
where, $\rho$, $p$, ${\bf v}$, ${\bf B}$, $\mu$, $\gamma$, and $k$ are density, pressure, flow velocity, magnetic field, magnetic permeability, atomicity coefficient, and a constant, respectively.
 For the derivation of a metric for magnetoacoustic wave we
 need to linearised the MHD equations with choosing the velocity disturbances as a
 velocity potential in similar analytical process for derivation of acoustic metric
 from wave equation. To do this,  first we treat
 the propagation of magnetoacoustic waves in homogenous unbounded medium choosing velocity disturbance
 as a velocity potential and a vector potential. Second, the magnetoacoustic and Alfv\'{e}nic  metrics are calculated.

\subsection{Helmholtz theorem and magnetoacoustic waves}  Properties of magnetoacoustic waves (fast, slow, and Alfv\'{e}n waves) in an unbounded  homogenous and stationary medium with constant density ($\rho_0=cte$), constant pressure ($p_0 =cte$), and uniform background magnetic field (${\bf B}_0$), were investigated in the literature (e.g., \citep{27}). The linearised ideal MHD equations can be reduced to a single equation for disturbed velocity as \citep{27},

\begin{eqnarray}
\frac{\partial ^2{\bf v}}{\partial t^2}=c_0^2\nabla(\nabla . {\bf v})+\text v_A^2(\nabla\times(\nabla\times({\bf v}\times {\hat B}_0)))\times {{{\hat B}_0}}.
\end{eqnarray}
 where $c_0=\sqrt{\frac{\gamma p_0}{\rho_0}}$ is the sound speed, $\text v_A={\frac{B_0}{\sqrt{\mu\rho_0}}}$ is speed of the Alfv\'{e}n wave, and ${\hat B_0}={\bf B}_0/{B_0}$ is a unit vector.

 Here, we focus on studying of the characteristics of the above mentioned waves using a fundamental theorem of calculus (Helmholtz's theorem). Based on the Helmholtz decomposition, a vector field with sufficient smoothness and decay conditions \citep{27-1}, can be decomposed to an irrotational part ($\nabla \phi$ where $\phi$ is a scaler) and a solenoidal part ($\nabla \times {\bf A}$ where ${\bf A}$ is the vector potential and satisfy $\nabla . {\bf A} =0$ ). The irrotational  (gradient) and divergence-free solutions can be used for treating the longitudinal and transversal waves, respectively.

  Suppose an irrotational solution, $\bf {v}=\nabla\varphi$ with $\varphi=\tilde{\varphi}\exp{(i{\bf k}.{\bf r}-i\omega t)}$ for Eq. (6), immediately we find  ${\bf v}=\hat { v} \exp{(i{\bf k}.{\bf r}-i\omega t)}$  where $\bf k$, $\omega$, $\tilde\varphi$, and $\hat{v}=i{\bf k}\tilde\varphi$, are wave vector, angular frequency of oscillations, a constant, and  wave amplitude, respectively. Furthermore Eq. (6) yields,
\begin{eqnarray}
\omega ^2{\hat{ v}}=c_0^2{\bf k}({\bf k} . {\hat{v}})+\text v_A^2({\bf k}\times({\bf k}\times({\hat { v}}\times {\hat B}_0)))\times {\hat B}_0.
\end{eqnarray}
 Equation (7) simplifies as
\begin{eqnarray}
\omega ^2\hat { v}=c_0^2 k^2\hat { v}+\text v_A^2(\hat { v}-{\hat B}_0(\hat { v}.{\hat B}_0))k^2.
\end{eqnarray}
We note that the wave vector ($\bf k$) and  wave amplitude ($\hat v$) are parallel (${\bf k}\|\hat {v}$).
Considering the wave propagation parallel to the background magnetic field  $({\bf k}\|{\bf B}_0)$, Eq. (8) reduces to the dispersion relation $\omega^2=c_0^2k^2$. This indicates that, in the case of an irrotational solution parallel to the magnetic field, only the acoustic wave can propagate. In the case of propagation perpendicular to the background magnetic field (${\bf k}\bot{\bf B}_0$), Eq. (8) gives the dispersion relation $\omega^2=({\text v_A^2+c_0^2}){k^2}$. This is the well-known characteristic of the fast magnetoacoustic wave with the  phase speed $c_f^2={\omega^2}/{k^2}=c_0^2+\text v_A^2 $. For the oblique propagation ${\bf k}.{\bf B}_0=kB_0\cos \theta~ (\theta$  is the angle between  ${{\bf B}_0}$ and ${\bf k})$, the phase speed of the slow  magnetoacoustic wave is given by
\begin{eqnarray}
v_{ph}^2=\frac{\omega^2}{k^2}=c_0^2-\text v_A^2(\cos \theta-1).
\end{eqnarray}

From the above analysis we see that, choosing the disturbed velocity as an irrotational velocity field $\bf v=\nabla\varphi$, the Alfv\'{e}n waves  cannot propagate along the background magnetic field (${\bf B}_0$).

Suppose a divergence-free solution, $\bf {v}=\nabla\times{\bf A}$, with a planar wave solution   ${\bf A}={\hat A}\exp{(i{\bf k}.{\bf r}-i\omega t)}$, $\hat{ v}=i{\bf k}\times\hat{A}$ and $\hat A$ is a constant vector, Eq. (6) gives,
\begin{eqnarray}
\omega ^2\hat{ v} =\text v_A^2({\bf k}\times({\bf k}\times(\hat{ v}\times {\hat B}_0)))\times {{\hat B}_0}.
\end{eqnarray}
We see that the velocity amplitude is perpendicular to the wave vector  (${\bf k}.{\hat v}=0$), and only the transversal Alfv\'{e}n wave with phase speed $v_{ph}=\text v_A$ can propagate.

In the remainder of this section, the metrics for acoustic, magnetoacoustic, and Alfv\'{e}n
longitudinal waves (irrotational solutions) are derived.
\subsection{Magnetoacoustic metric}

Here, in the presence of initial material flow the magnetoacoustic metric using the magnetoacoustic wave is derived. For an irrotational flow the velocity is satisfied by a scalar field ${\bf v}=\nabla\varphi$. We consider small perturbations from equilibrium as
\begin{eqnarray}\label{pert}
\rho({\bf r},t)&=&\rho_0(x,t)+\rho_0(x,t)\psi({\bf r},t),\label{p1}\\
{\bf v}({\bf r},t)&=&\text {v}_0(x)\hat{x}+\nabla\varphi({\bf r},t),\\
{\bf B}({\bf r},t)&=&{B}_0\hat{x}+{\bf B}_1({\bf
r},t)\label{p3},
\end{eqnarray}
where, equilibrium quantities indicated by subscript "0" are function of position $x$ and time $t$, $\psi({\bf r},t)$, $\varphi({\bf r},t)$, and ${\bf B}_1({\bf r},t)$ are perturbed quantities \citep{28, 29}.
Equilibrium quantities ($\rho_0, \text v_0, {\bf B}_0,$ and~~$ p_0$) are satisfied by
 \begin{eqnarray}
&&\frac{{\partial \rho_0 }} {{\partial t}} +\text v_0\frac{{\partial \rho_0 }} {{\partial x}}+\rho_0\frac{{\partial \text v_0 }} {{\partial x}}=0,\label{cont0}\\
 &&
 \rho_0 \text v_0\frac{\partial { \text v_0} }{{\partial x}}= -c_0^2\frac{{\partial \rho_0 }} {{\partial x}}\label{momen0}\\
&&\nabla\times({\bf
\text {\bf v}}_0\times{\bf B}_0)=0,\label{induc0}\\
&&\nabla.{\bf B}_0=0,\label{solo}\\
&&p_0=k\rho_0^\gamma.\label{adia0}
\end{eqnarray}

 Linearization of Eqs (\ref{cont})-(\ref{adia}) (products and squares of the small perturbations are neglected) and after some mathematical manipulations, give
\begin{eqnarray}
  \frac{{\partial \psi }}
{{\partial t}} + \nabla {\text{}}\ln {\rho _0}.\nabla \varphi  + {{\mathbf{v}}_0}.\nabla \psi  + {\nabla ^2}\varphi  = 0,\,\,\,\,\,\,\,\,\,\,\,\,\,\,\,\,\,\,\,\,\,\,\,\,\,\,\,\,\,\,\, \hfill \label{19}\\
  {\rho _0}\nabla \left( {\frac{{\partial \varphi }}
{{\partial t}} + {{\mathbf{v}}_0}.\nabla \varphi  + c_0^2\psi } \right) = \frac{1}
{\mu }\left( {\nabla  \times {{\mathbf{B}}_1}} \right) \times {{\mathbf{B}}_0},\,\,\,\,\, \hfill \label{20}\\
  \frac{{\partial {{\mathbf{B}}_1}}}
{{\partial t}} = \nabla  \times \left( {{{\mathbf{v}}_0} \times {{\mathbf{B}}_1}{\text{}}} \right) + \nabla  \times \left( {{{\mathbf{v}}_1} \times {{\mathbf{B}}_0}{\text{}}} \right),{\text{}}\,\,\,\,\,\,\,\,\,\,\,\,\,\,\,\,\,\,\,\,\,\,\, \hfill \label{21}\\
  \nabla .\,{{\mathbf{B}}_1} = 0,\,\,\,\,\,\,\,\,\,\,\,\,\,\,\,\,\,\,\,\,\,\,\,\,\,\,\,\,\,\,\,\,\,\,\,\,\,\,\,\,\,\,\,\,\,\,\,\,\,\,\,\,\,\,\,\,\,\,\,\,\,\,\,\,\,\,\,\,\,\,\,\,\,\,\,\,\,\,\,\,\,\,\,\,\,\,\,\,\,\,\,\,\,\,\,\,\,\,\,\,\,\,\,\label{22}
\end{eqnarray}
As we explained in the previous section, by choosing ${\bf v}=\nabla \varphi$ the parallel propagation of Alfv\'{e}n waves along the magnetic field $({\bf B}_0=B_0 \hat x)$ is absent. Therefore, the first term on right hand of  Eq. (\ref{21}) is set to zero, $\nabla  \times \left( {\mathbf{v}}_0 \times {\mathbf{B}}_1 \right) = 0$. A combination of this assumption and the solenoidal condition for magnetic field $(\nabla.{\bf B}_0=0)$, gives $\frac{\partial B_{1x}}{\partial x}=0$. Thus the propagation of Alfv\'{e}n waves along the background magnetic field is removed from our analysis.
We assume the irrotational part of the vector ${{{\mathbf{v}}_0} \times {{\mathbf{B}}_1}}=\nabla \delta(x,y,z,t)$ in which $\delta(x,y,z,t)$ is a function.

 The $x$ and $z$ component of Eq. (\ref{20}) are
 \begin{eqnarray}\label{24}
 \frac{\partial }{\partial x} \left( {\frac{{\partial \varphi }}
{{\partial t}} + {{\mathbf{v}}_0}.\nabla \varphi  + c_0^2\psi } \right) = 0.
 \end{eqnarray}
 \begin{eqnarray}\label{23}
 {\rho _0}\frac{\partial }{\partial z} \left( {\frac{{\partial \varphi }}
{{\partial t}} + {{\mathbf{v}}_0}.\nabla \varphi  + c_0^2\psi } \right) = \frac{{{B}}_0}
{\mu }\left( \frac{\partial B_z}{\partial x}-\frac{\partial B_x}{\partial z} \right).
 \end{eqnarray}

 Here after, first, we focus our analysis on the derivation of acoustic waves with propagation along the background magnetic field directions $(\hat x)$ and second, magnetoacoustic wave propagation in all directions
 except the magnetic field direction.

 First, combining Eqs (\ref{19}) and (\ref{24}) gives
\begin{eqnarray}\label{K-G}
   - \frac{\partial }
{{\partial t}}\left( {\frac{{\partial \varphi }}
{{\partial t}} + {{\text{v}}_0}\frac{{\partial \varphi }}
{{\partial x}}} \right) + \frac{\partial }
{{\partial x}}\left( {c_{0}^2\frac{{\partial \varphi }}
{{\partial x}} - {{\text{v}}_0}\left( {\frac{{\partial \varphi }}
{{\partial t}} + {{\text{v}}_0}\frac{{\partial \varphi }}
{{\partial x}}} \right)} \right)
  + c_{0}^2\nabla^2 \varphi = 0.
\end{eqnarray}
Equation (\ref{K-G}) is the well-known Klein-Gordon equation for acoustic waves.
 Eliminating
$\partial \psi /\partial x$ between Eq. (\ref{19}) and Eq. (\ref{24}) one obtains
\begin{eqnarray}\label{26}
&&\text{v}_0\frac{\partial }
{\partial x}\left( {\frac{{\partial \varphi }}
{{\partial t}} + {{\mathbf{v}}_0}.\nabla \varphi } \right) + \text{v}_0\psi \frac{{\partial c_0^2}}
{{\partial x}}\nonumber\\&&~~~ - c_0^2\left(\frac{\partial \psi}
{\partial t} + \nabla \ln {\rho _0}.\nabla \varphi  + \nabla ^2\varphi\right) = 0\,.\,\,\,\,\,\,
\end{eqnarray}
Briefly, by choosing the irrotational solution for the velocity disturbance, the transversal Alfv\'{e}n wave is absent, and the propagation of the longitudinal Alfv\'{e}n wave parallel with the background magnetic field is also absent  as expected. Our analysis shows that, along the magnetic field only the acoustic wave can  propagate.
Because our goal is to analyse the magnetoacoustic black hole, hereafter, we focus our analysis in all directions except the magnetic field direction.

Second, by differentiating Eq. (\ref{26}) with respect to $z$ and substituting $\partial \psi /\partial z$ from Eq. (\ref{23}) one finds
\begin{eqnarray}
  \frac{\partial }
{{\partial z}}\left( {\frac{1}
{{c_0^2}}\frac{\partial }
{{\partial x}}\frac{{d\varphi }}
{{dt}}} \right) = \frac{1}
{{{\text v_0}}}{\nabla ^2}\frac{{\partial \varphi }}
{{\partial z}} + \frac{1}
{{{\text v_0}}}\nabla {\text{}}\ln {\rho _0}.\nabla \frac{{\partial \varphi }}
{{\partial z}} \hfill \nonumber\\
\,\,\,\,\,\,\,\,\,\,\,\,\,\,\,\,\,\,\, + \frac{1}
{{{\text v_0}}}\frac{\partial }
{{\partial t}}\left( { - \frac{{{B_0}}}
{{\mu {\rho _0}c_{0}^2}}{{\left( {\nabla  \times {{\mathbf{B}}_1}} \right)}_y} - \frac{1}
{{c_0^2}}\frac{\partial }
{{\partial z}}\frac{{d\varphi }}
{{dt}}} \right) \hfill \nonumber\\
\,\,\,\,\,\,\,\,\,\,\,\,\,\,\,\,\, + \frac{{{B_0}}}
{{\mu {\rho _0}c_0^4}}\frac{{\partial c_{0}^2}}
{{\partial x}}{\left( {\nabla  \times {{\mathbf{B}}_1}} \right)_y} + \frac{\partial }
{{\partial z}}\left( {\frac{1}
{{c_0^4}}\frac{{\partial c_{0}^2}}
{{\partial x}}\frac{{d\varphi }}
{{dt}}} \right).{\text{      }} \hfill  \label{27}
\end{eqnarray}
in which, $d/dt = \partial /\partial t + {{\mathbf{v}}_0}.\nabla$. Substituting
$- \frac{\partial }
{{\partial t}}{\left( {\nabla  \times {{\mathbf{B}}_1}} \right)_y} = {B_0}{\nabla ^2}\frac{{\partial \varphi }}
{{\partial z}}$  from Eq. (\ref{21}), into Eq. (\ref{27}) and differentiating  with respect to $t$ using $\frac{\partial }
{{\partial t}}{\left( {\nabla  \times {{\mathbf{B}}_1}} \right)_y} = -{B_0}{\nabla ^2}\frac{{\partial \varphi }}
{{\partial z}}$  we derived a single equation  for velocity potential $(\varphi)$
\begin{eqnarray}\label{28}
  \frac{\partial }
{{\partial t}}\left( {\frac{1}
{N}\left( {\frac{1}
{{c_0^2}}\frac{\partial }
{{\partial x}}\frac{{d\varphi }}
{{dt}} - \frac{1}
{{c_0^4}}\frac{{\partial c_{0}^2}}
{{\partial x}}\frac{{d\varphi }}
{{dt}} - \frac{{{\nabla ^2}\varphi }}
{{{{\text{v}}_0}}} - \frac{{\nabla {\text{ln}}{\rho _0}.\nabla \varphi }}
{{{{\text{v}}_0}}}} \right)} \right) \hfill \nonumber\\
   + \frac{\partial }
{{\partial t}}\left( {\frac{1}
{N}\left( {\frac{1}
{{{{\text{v}}_0}}}\frac{\partial }
{{\partial t}}\left( {\frac{1}
{{c_0^2}}\frac{{d\varphi }}
{{dt}}} \right) - \frac{{{\text{v}}_{{\text{A}}}^2}}
{{c_{0}^2{{\text{v}}_0}}}{\nabla ^2}\varphi } \right)} \right) = {B_0}{\nabla ^2}\varphi {\text{, }}
\,\,{\text{    }} \hfill
\end{eqnarray}
where,
\[\begin{gathered}
  N = \frac{{{B_0}}}
{{\mu {{\text{v}}_0}}}\frac{\partial }
{{\partial t}}\left( {\frac{1}
{{{\rho _0}c_{0}^2}}} \right) - \frac{{{B_0}}}
{{\mu {\rho _0}c_{0}^4}}\frac{{\partial c_{0}^2}}
{{\partial x}} \hfill \\
  \,\,\,\,\,\,\, = \frac{{{{\text{v}}^2_A}}}
{{{B_0}c_{0}^2}}\,\left( {\frac{\gamma }
{{{\text{v}_0}}} + \frac{{{\text{v}_0}}}
{{c_0^2}}} \right)\,\frac{{d{\text{v}_0}}}
{{dx}}. \hfill \\
   \hfill \\
\end{gathered} \]
For a flow having a slight change in the speed $(d{{\text{v}}_0}/dx\ll1)$, for high frequency waves (short period $\Delta t\ll1$)  the term $\Delta t(d{{\text{v}}_0}/dx)$ becomes too small. In this case, the right hand side term ${B_0}{\nabla ^2}\varphi$ of Eq. (\ref{28}) can be negligible compared to the last term $\frac{\partial }
{{\partial t}} \left({\frac{{{\text{v}}_{{\text{A}}}^2}}
{{N c_{0}^2{{\text{v}}_0}}}{\nabla ^2}\varphi }\right)$ of the left hand side.  This leaves an equation for $\varphi$
\begin{eqnarray}\label{29}
   - \frac{\partial }
{{\partial t}}\left( {\frac{{\partial \varphi }}
{{\partial t}} + {{\text{v}}_0}\frac{{\partial \varphi }}
{{\partial x}}} \right) + \frac{\partial }
{{\partial x}}\left( {c_{0}^2\frac{{\partial \varphi }}
{{\partial x}} - {{\text{v}}_0}\left( {\frac{{\partial \varphi }}
{{\partial t}} + {{\text{v}}_0}\frac{{\partial \varphi }}
{{\partial x}}} \right)} \right) \hfill \nonumber\\
  \,\,\,\,\,\,\,\,\,\,\,\,\,\, +{\text{v}}_{{\text{A}}}^2\frac{{{\partial ^2}\varphi }}
{{\partial {x^2}}} + {\text{(v}}_{{\text{A}}}^2 + c_{0}^2)\left( {\frac{{{\partial ^2}\varphi }}
{{\partial {y^2}}} + \frac{{{\partial ^2}\varphi }}
{{\partial {z^2}}}} \right) = 0.\,\,\,\,\,\, \hfill
\end{eqnarray}
Equation (\ref{29})  describes the propagation of acoustic, Alfv\'{e}n, and magnetoacoustic waves in laboratory and astrophysical plasma. This equation is in the form of the well-known Klein-Gordon equation. As expected, in the case of unmagnetized fluid ($B_0=0$), Eq. (\ref{29}) reduces to the acoustic wave equation for velocity potential. Usually, a d'Alembertian equation (for a minimally coupled massless scalar field) of motion was derived for velocity potential in a barotropic, inviscid, and rotational free flow \citep{25}.

  Equation (\ref{29}) can be reformulate in a tensorial form
  \begin{eqnarray}
\frac{1}
{{\sqrt { - g} }}{\partial _\mu }\left( {\sqrt { - g} {g^{\mu \nu }}{\partial _\nu }\varphi } \right) = 0,\,\,\,\,\,\,\,\,\,\,\,\,\,\,\,\,\,\,\,\,\,\,\,\,\,\,\,\,\,\,\,\,\,\,\,
\end{eqnarray}
where,  $g^{\mu \nu }$ and $g$ are inverse metric tensor and its determinant,  $\mu$ and $\nu$  runs from 0 (indicates the time coordinate) to 3 (denotes the spatial coordinate). The magnetoacoustic inverse metric tensor $g^{\mu \nu }$  and metric tensor  $g_{\mu \nu }$ are obtained as
\begin{eqnarray}\label{31}
  {g^{\mu \nu }} = \frac{1}
{{{\rho _0}c_{f}^3}}\left( {\begin{array}{*{20}{c}}
   { - 1} & { - {\text{}}{{\text{v}}_0}} & 0 & 0  \\
   { - {\text{}}{{\text{v}}_0}} & {c_f^2 - {\text{v}}_{0{\text{}}}^2} & 0 & 0  \\
   0 & 0 & {c_f^2} & 0  \\
   0 & 0 & 0 & {c_f^2}  \\

 \end{array} } \right),\,\,\,\,\, \hfill
  {g_{\mu \nu }} =  - {\rho _0}{c_f}\left( {\begin{array}{*{20}{c}}
   {c_f^2 - {\text{v}}_{0{\text{}}}^2} & {{\text{}}{{\text{v}}_0}} & 0 & 0  \\
   {{\text{}}{{\text{v}}_0}} & { - 1} & 0 & 0  \\
   0 & 0 & { - 1} & 0  \\
   0 & 0 & 0 & { - 1}  \\

 \end{array} } \right),\,\,\,\,\,\,\,\,\, \hfill
\end{eqnarray}
where, $c_{f}^2 = {\text{v}}_{{\text{A}}}^2 + c_{0}^2$. Using metric tensor, Eq. (\ref{31}), the magnetoacoustic interval can be defined as
\begin{eqnarray}\label{32}
  d{s^2} = {g_{\mu \nu }}d{x^\mu }d{x^\nu } \hfill
   =  - {\rho _0}{c_f}\left( {(c_{f}^2 - {\text{v}}_0^2)d{t^2} + 2{{\text{v}}_0}dtdx - d{x^2} - d{y^2} - d{z^2}} \right).\,\, \hfill
\end{eqnarray}
Inserting the specific time interval $d\tau  = dt + \frac{{{{\text{v}}_{\text{0}}}dx}}
{{c_{f}^2 - {\text{v}}_{0{\text{}}}^2}}$  into Eq. (\ref{32}), one obtains
\begin{eqnarray}\label{33}
  d{s^2} =  \hfill
  {c_f}{\rho _0}\left( { - \left( {1 - \frac{{{\text{v}}_{0{\text{}}}^2}}
{{c_{f}^2}}} \right)c_{f}^2d{\tau ^2} + \frac{d{x^2}}{{{\left( {1 - \frac{{{\text{v}}_{0{\text{}}}^2}}
{{c_{f}^2}}} \right)}}} + d{y^2} + d{z^2}} \right). \hfill
\end{eqnarray}
In the following section, the properties of acoustic, magnetoacoustic, and Alfv\'{e}nic BHs are investigated.

\section{Acoustic, Magnetoacoustic, and Aflv\'{e}nic  black holes}\label{sec: bhs}
\subsection{Acoustic black hole}
The theory of gravitational BHs has been developed into the supersonic flow by Unruh \citep{26}.
 For a moving fluid medium at the horizon where speed of medium is closed to propagation speed of
 the acoustic signals " then nothing can fight its way back upstream and signals are trapped" \citep{31}.
 In the case of unmagnetised gas($B_0=0$), Eq. (\ref{28}) reduces to the Klein-Gordon equation, for propagation of acoustic waves in the presence of material flow. Using the resultant equation the acoustic metric can be derived \citep{26}.

\begin{figure}
\centerline{\includegraphics[width=2\columnwidth]{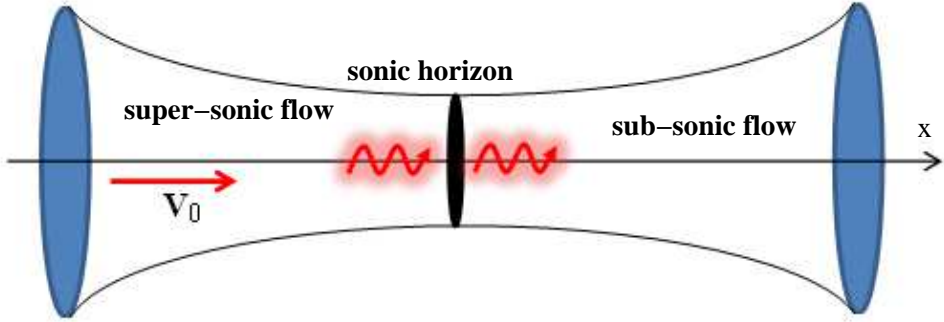}}
\caption{A schematic presentation of a acoustic black hole in Laval  nozzle tube.
  The speed of super sonic flow  reduces to local propagation speed of acoustic wave at sonic horizon; then acoustic signals are trapped.}
  \label{fig 1}
\end{figure}

 The acoustic metric can be obtained by setting $\text v_A=0$ in Eq. (\ref{31})
 \begin{eqnarray}\label{42}
g_{\mu \nu }^s =  - {\rho _0}{\text c_0}\left( {\begin{array}{*{20}{c}}
   {\text c_0\texttt{}^2 - {\text{v}}_{0{\text{}}}^2} & {{\text{}}{{\text{v}}_0}} & 0 & 0  \\
   {{\text{}}{{\text{v}}_0}} & { - 1} & 0 & 0  \\
   0 & 0 & { - 1} & 0  \\
   0 & 0 & 0 & { - 1}  \\

 \end{array} } \right).\,\,\,\,\,\,\,\,\,\,\,\,\,\,\,\,\,\,\,\,\,\,\,\,\,\,\,\,\,
 \end{eqnarray}
Equivalently, the acoustic interval can be expressed as
  \begin{eqnarray}\label{43}
  d{s^2} =  \hfill
  {c_0}{\rho _0}\left( { - \left( {1 - \frac{{{\text{v}}_{0{\text{}}}^2}}
{{c_{0}^2}}} \right)c_{f}^2d{\tau ^2} + \frac{d{x^2}}{{{\left( {1 - \frac{{{\text{v}}_{0{\text{}}}^2}}
{{c_{0}^2}}} \right)}}} + d{y^2} + d{z^2}} \right). \hfill
\end{eqnarray}
Combination of continuity and momentum equations (Eqs \ref{cont0} and \ref{momen0}) in stationary state, the relation between cross section $S$ and velocity $\text{v}$  is given by
\begin{eqnarray}\label{34}
\left( {\frac{{{\text{v}}_{0{\text{}}}^2}}
{{c_{0}^2}} - 1} \right)\frac{{d{{\text{v}}_0}}}
{{{{\text{v}}_0}}} = \frac{{dS}}
{S}{\text{.      }}
\end{eqnarray}
This relation shows that for $(dS < 0)$ a subsonic flow $({{\text{v}}_0} < {c_0})$ will be accelerated and a supersonic flow  $({{\text{v}}_0} > {c_0})$ will be decelerated.
If the nozzle is sufficiently narrow  and with a slightly variable cross section the speed of flow exceeds to sound speed at the throat (sonic horizon). This shows the acoustic interval Eq. (\ref {43}) interpreted acoustic BH which has a sonic horizon and trapped phonon in the surface gravity of acoustic BH. This means that,  when the acoustic waves cross from upstream to downstream, the acoustic wave quanta (phonons)  are captured in the horizon of the BH where they emit Hawking radiation made by phonons. In this regard, in many papers a Laval nozzle setup Figure \ref{fig 1} has been proposed to discuss the above mentioned acoustic BH. This setup uses, an axisymmetric sufficiently thin tube with slightly decreasing cross section $\left( {S(x)} \right)$ that reaches its minimum cross section at the throat and then slightly increases that. An initial material flow $({{{\bf v}}_0} = {{\text{v}}_0}\left( x \right)\,\hat x)$ along the tube axis is considered.

\subsection{Magnetoacoustic black hole}
\begin{figure}
\includegraphics[width=1.1\columnwidth]{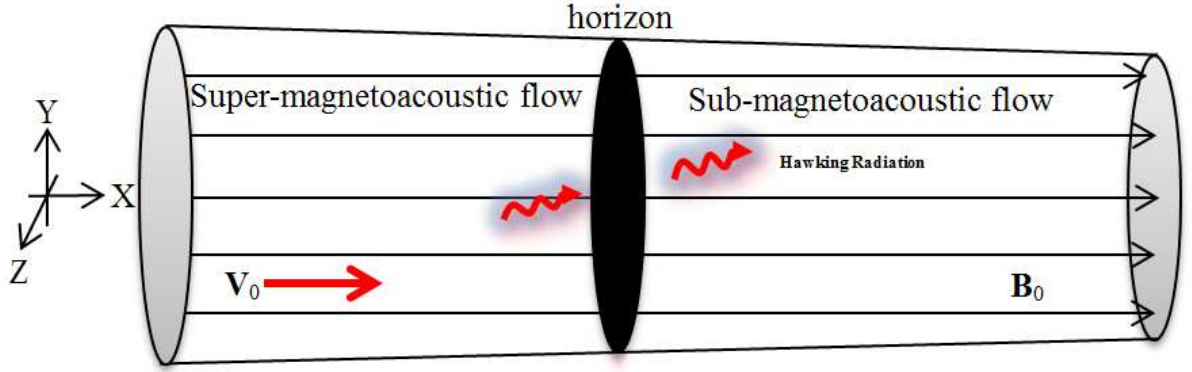}
\caption{Sketch of a tube with a slightly variable cross section
embedded in a slightly uniform magnetic field. Super-magnetoacoustic
flow, $\text v_0 > c_f $, sub-magnetoacoustic flow,
$\text v_0 < c_f$ , and magnetoacoustic wave trapped in the horizon
are presented.}
\label{fig 2}
\end{figure}

The magnetoacoustic metric Eq. (\ref{33}) is singular at the magnetoacoustic point,
where ${\text{}}{{\text{v}}_0} = {c_f}$, determines a magnetoacoustic horizon. The speed of super magnetoacoustic plasma flow  reduces to local propagation speed of magnetoacoustic wave at horizon; then signal of magnetoacoustic wave is trapped and therefore it can be called magnetoacoustic BH. Similar to  the HR emitted from acoustic and gravitational BHs, the magnetoacoustic BH also should emit HR. In this regard, we propose a setup Figure \ref{fig 2} to discuss the above mentioned BH. The setup consists of an axisymmetric sufficiently thin tube with slightly variable cross section $\left( {S(x)} \right)$, a uniform force free magnetic field  ${\bf B}_0 = {{\text{B}}_0}\hat x$ and an initial material flow $({{{\bf v}}_0} = {{\text{v}}_0}\left( x \right)\,\hat x)$ along the tube axis Figure \ref{fig 2}.

A similar treatment of sub and supersonic flow in tube configuration can be explained for sub and super-magnetoacoustic flow
based on Eq. (\ref{34}). In other words, the super-magnetoacoustic flow $({{\text{v}}_0} > {c_f})$  will be decelerated along the tube where its cross section slightly decreases. It is possible to release a super-magnetoacoustic flow in the tube, which its speed tends to the speed of magnetoacoustic wave (${{\text{v}}_0}={c_f}$)  at the horizon. The boundary between sub-magnetoacoustic and super-magnetoacoustic flow could be called the magnetoacoustic horizon, analogous to the sonic horizon in acoustic BHs. Phononic version of HR is an inevitable result of trapping acoustic wave at the acoustic horizon \citep{24,30}. Under a likely scenario, the magnetoacoustic wave cannot escape from the magnetoacoustic horizon, therefore should emit HR made of magnephonon. Indeed, a magnephonon will be a quantum for magnetoacoustic wave, analogous to the phonon which is a quantum for acoustic wave.

\subsection{Alfv\'{e}nic black hole}
In the limit of  ${{\text{c}}_0} << {{\text{v}}_{\text{A}}}$ (zero $\beta$ plasma condition), the magnetoacoustic wave (Eq. \ref{29}) reduces to an Alfv\'{e}n wave in the presence of initial material flow. The resultant Alfv\'{e}n wave equation is in the form of Klein-Gordon equation. Immediately, the Alfv\'{e}nic metric can be derived from magnetoacoustic metric (Eqs \ref{31} and \ref{33}) by setting $c_0$ tends to 0,
\begin{eqnarray}\label{35}
g_{\mu \nu }^A =  - {\rho _0}{\text v_A}\left( {\begin{array}{*{20}{c}}
   {\text v_A^2 - {\text{v}}_{0{\text{}}}^2} & {{\text{}}{{\text{v}}_0}} & 0 & 0  \\
   {{\text{}}{{\text{v}}_0}} & { - 1} & 0 & 0  \\
   0 & 0 & { - 1} & 0  \\
   0 & 0 & 0 & { - 1}  \\

 \end{array} } \right).\,\,\,\,\,\,\,\,\,\,\,\,\,\,\,\,\,\,\,\,\,\,\,\,\,\,\,\,\,
 \end{eqnarray}

\begin{figure}
\includegraphics[width=1.1\columnwidth]{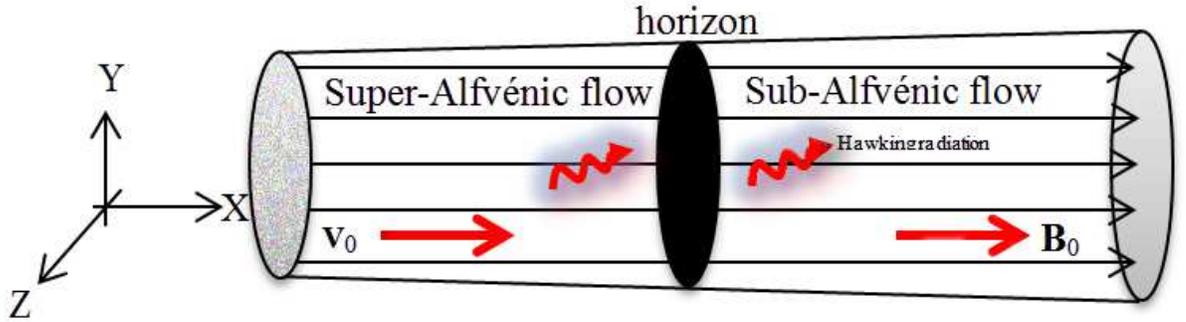}
\caption{A schematic presentation of a Alv\'{e}nic black hole.  Sketch of a tube with a slightly variable cross section embedded in a uniform magnetic field.
  The speed of super Alfv\'{e}nic flow  reduces to local propagation speed of Alfv\'{e}nic wave at horizon; then Alfv\'{e}nic signals are trapped.
} \label{fig 3}
\end{figure}

Equivalent Alfv\'{e}nic interval is given by
\begin{eqnarray}\label{36}
  d{s^2} =  \hfill
  {\text v_A}{\rho _0}\left( { - \left( {1 - \frac{{{\text{v}}_{0{\text{}}}^2}}
{{\text v_{A}^2}}} \right)\text v_{A}^2d{\tau ^2} +\frac{d{x^2}}{{{\left( {1 - \frac{{{\text{v}}_{0{\text{}}}^2}}
{{\text v_{A}^2}}} \right)}}}  + d{y^2} + d{z^2}} \right){\text{.}} \hfill
\end{eqnarray}
The Alfv\'{e}nic metric (Eq. \ref{36}) is singular at the location of Alfv\'{e}nic point where ${{\text{v}}_0} = {{\text{v}}_{\text{A}}}$. This singular behaviour of Alfv\'{e}nic metric leads to an Alfv\'{e}nic BH. We propose a setups that is illustrated in Figure \ref{fig 3}, the condition for occurrence of an Alfv\'{e}nic BH can be discussed. For compressional (longitudinal) Alfv\'{e}n wave, the plasma density is nearly constant. As a result of mass continuity, the flux, $S(x){\text{v}_0}(x)$, is  constant in the tube cross section. Consider a super-Alfv\'{e}nic flow in a tube with increasing cross section Figure \ref{fig 3}, speed of the flow decreases along the tube axis and reaches to Alfv\'{e}n velocity $\text{v}_0=\text{v}_A$ at Alfv\'{e}nic point and then transformed to sub-Alfv\'{e}nic flow. Therefore, we call Alfv\'{e}nic horizon to be the interface between super-Alfv\'{e}nic flow and sub-Alfv\'{e}nic flow.  In horizon of the Alfv\'{e}nic BH, the Alfv\'{e}n wave is trapped and it is expected to be radiated by Alphonon. Alphonon is introduced as a quantum particle for Alfv\'{e}n wave packet.


\section{Hawking temperature}\label{sec: HT}
 Hawking temperature is an important characteristic of BHs. Unruh showed that fluid flows mimic BHs. Hawking temperature ($T_H$) for acoustic BH was obtained
\begin{eqnarray}\label{37}
{T_{\rm H - acoustic}} = \frac{\hbar }
{{2\pi k{c_s}}}{\left. {\frac{{d({\text{v}}_0^2 - c_s^2)}}
{{dx}}} \right|_{{\text{horizon}}}}{\text{, }}\,\,\,\,{\text{}}\,{\text{}}
\end{eqnarray}
where, $\hbar=h/2\pi$, $h$  is the plank constant and $k$  is the Boltzmann constant.  Since the Hawking temperature is independent of metric conformal factor. It will therefore be as following for the magnetoacoustic BH,
\begin{eqnarray}\label{38}
{T_{\rm H - magnetoacoustic}} = \frac{\hbar }
{{2\pi k{c_f}}}{\left. {\frac{{d({\text{v}}_0^2 - c_f^2)}}
{{dx}}} \right|_{{\text{horizon}}}}{\text{.}}\,\,\,\,\,\,\,{\text{}}\,{\text{}}
\end{eqnarray}
In the limit of  $\text{v}_A$ tends to zero, Eq. (\ref{38}) then reduces to Hawking temperature for acoustic BH.
Although, in the zero $\beta$ plasma condition the above mentioned equation can describe the Hawking temperature for Alfv\'{e}nic BH,
\begin{eqnarray}\label{39}
{T_{\rm H -\text{ Alfv\'{e}nic}}} = \frac{\hbar }
{{2\pi k{{\text{v}}_A}}}{\left. {\frac{{d({\text{v}}_0^2 - {\text{v}}_A^2)}}
{{dx}}} \right|_{{\text{horizon}}}}{\text{.}}\,\,\,\,\,\,\,{\text{}}\,{\text{}}
\end{eqnarray}
At the horizon where the tube cross section radius is equal to $R$, Eq. (\ref{38}) can be simplified as
\begin{eqnarray}\label{40}
{T_{H - magnetoacoustic}} \approx \frac{{\hbar \,{{\text{c}}_f}}}
{{2\pi k\,R}},\,\,\,\,\,\,\,\,\,\,\,\,\,\,\,\,\,\,\,\,\,\,\,\,\,\,\,\,\,\,\,\,\,\,\,\,\,
\end{eqnarray}
where, the term $\frac{1}
{{{c_f}}}\frac{d}
{{dx}}({c_f} - {{\text{v}}_0})$ is approximately equal to $\frac{1}
{R}.$ For a plasma with a ratio of $\chi  = {c_0}/{{\text{v}}_A}$ in which $\chi$ is a positive number, Eq. (\ref{40}) gives
\begin{eqnarray}\label{41}
{T_{H - magnetoacoustic}} \approx 2.66 \times {10^4}{\left( {1 + {\chi ^2}} \right)^{0.5}}\frac{{{B_0}\,}}
{{R\sqrt n }},\,\,\,\,\,\,\,\,\,\,\,\,\,\,
\end{eqnarray}
where, $n$ is the number density of plasma and all units are in SI. For a typical plasma with magnetic field strength  $B_0$=1 Tesla, number density $n = {10^{18}}\,{m^{ - 3}}$, $R=1$ mm, and ${\chi ^2} \ll 1$, the Hawking temperature is estimated about 0.0266 K. For a typical natural fluid in a Lavel nozzle experiment, the Hawking temperature was estimated about $10^{-6}$ K \citep{31}.

\section{conclusion}\label{sec: conc}
In this study, the behavior of MHD waves (Alfv\'{e}n and magnetoacoustic) are are treated in the Helholtz decomposition frame-work, in the case of irrotational flow,  we introduced the magnetoacoustic and Alfv\'{e}nic analogue BHs. In the horizon of magnetoacoustic and Alfv\'{e}nic BHs, the magnetoacoustic and Alfv\'{e}n waves are trapped, respectively, and should emit magnephonons and Alphonons version of HR at Hawking temperature. The next logical step is to investigate the physical properties of both magnephonon and Alphonon quasi-particles based on quantum approach.
As stated in the literature, for acoustic BH in a natural fluid, the Hawking temperature is a function of sound speed and geometry of nozzle setup at the horizon. However, Hawking temperature for magnetoacoustic BH is related to the sound speed with additional positive terms  that depends on the magnetic field and density. Magnephonons and Alphonons are particles (quanta) corresponding to magnetoacoustic and Alfv\'{e}n waves, respectively.
The idea for definition of acoustic, magnetoacoustic, and Alfv\'{e}nic BHs can be applied to theoretical and/or experimental prediction of new non-gravitational BHs. Perhaps, the study of new BHs could help us observe HR.


\listofchanges
\end{document}